\documentclass[a4paper,11pt]{article}
\pdfoutput=1 

\usepackage{jheppub} 

\usepackage[T1]{fontenc} 

\usepackage{pdfsync}
\usepackage{mathptmx}
\usepackage[utf8]{inputenc}
\usepackage[T1]{fontenc}
\usepackage{slashed}
\usepackage{times}
\usepackage{amssymb,amsfonts,amsmath,amsthm}
\usepackage{dsfont,bbm}
\usepackage{dcolumn}
\usepackage{epsf}
\usepackage{graphicx}
\usepackage[caption=false]{subfig}
\usepackage{dsfont}
\usepackage{bm}
\usepackage{eucal}
\usepackage{slashed}
\usepackage[active]{srcltx}
\usepackage[usenames]{color}

\title{\boldmath Conformal Symmetry and Effective Potential: I.
Vacuum $V_{z,x}$-operation for the Green functions}

\author[a]{I.~V.~Anikin}

\affiliation[a]{Bogoliubov Laboratory of Theoretical Physics JINR, 141980 Dubna, Russia}

\emailAdd{anikin@theor.jinr.ru}

\abstract{We begin a series of two papers that is devoted to the study of the multi-loop effective potential
evolution in $\varphi^4$-theory using the  conformal symmetry.
In the first part, we introduce and describe in detail the vacuum $V_{z,x}$-operation
($``V"$ stems from  ``vacuum'', $\{z,x\}$ imply the
corresponding coordinates) that transforms
the given Green functions to the corresponding vacuum integrations which generate the effective potential.
Our operation can be considered as an inverse procedure compared to the Gorishni-Isaev method.
To the final goal, it is necessary to introduce also the special treatment of the mass terms as sorts
of ``interaction'' in an asymptotical expansion of the generating functional.}

\begin{document}
\maketitle
\flushbottom

\section{Introduction}

The effective potential approaches play an important
role for the different theoretical studies where the manifestations of spontaneous symmetry breaking are under
the main considerations. The special attentions are paid for the quantum corrections
which usually distort the classical geometrical picture related to the Goldstone theorem.
However, the presence of massive parameters (or particle masses) in the theory can significantly
complicate the multi-loop calculations even in the vacuum integrations which form the effective potential.
On the other hand, the massless propagators in the corresponding loops simplify any
multi-loop calculations and, moreover, open the window for the appropriate use of conformal symmetry \cite{Anikin2023}.

In the paper, based on the stationary phase method,
we use an alternative representation for the generating functional in $\varphi^4$-theory
where the massive term of Lagrangian has been treated as a sort of "effective interactions".
As a result of this, the scalar propagators in the vacuum diagrams describing interactions become massless ones \cite{Anikin2023}.

Due to the singular parts of the loop corrections, the effective potential demands the renormalization resulting in
the renormalization scale $\mu$ dependence. As usual, the scale evolution of effective potential is governed by the
anomalous dimension within the RG-method. It turns out that the anomalous dimension of effective potential
can be readily derived from the anomalous dimension of the non-local operator Green function
with the help of $V_{z,x}$-operation at any loop accuracy \cite{Anikin2023}.
In other words, if the anomalous dimension of the corresponding Green function is known
at the given $\ell$-loop accuracy (see for example \cite{Braun:2013tva}),
the anomalous dimension of effective potential formed by the vacuum integrations can be almost algebraically
calculated at the $(\ell +1)$-loop accuracy thanks for the new-introduced $V_{z,x}$-operation.
Generally speaking, the $V_{z,x}$-operation can be assumed as an inverse operation compared to
the Gorishni-Isaev method \cite{Gorishnii:1984te}, inspired by  \cite{Kazakov:1983pk, Kazakov:1983ns},
where the Green functions of propagator-type have been reduced to
the vacuum integration. The findings of \cite{Gorishnii:1984te}  became extremely useful for the multi-loop calculations.

In the paper, we give an comprehensive description of $V_{z,x}$-operation which is one of the main tool used in
the extended version  \cite{Anikin2023}.

\section{The masslessness procedure of Effective Potential in $\varphi^4$}
\label{bsec:GenFun}

We begin with the generating functional in the scalar $\varphi^4$ theory which leads to the effective action/potential
\footnote{Since, as well-known, the effective action differs from the effective potential by the (infinite) space-time volume,
$V\times T\sim \delta^{(4)}(0)$, we neglect this difference unless it leads to misunderstanding.}.
In a theory with massive parameters and interactions, the generating functional has the following form
(modulo the normalization constants denoted as {\it n.c.}):
\begin{eqnarray}
\label{GenFun-1}
&&{\mathbb Z}[J]\stackrel{n.c.}{=}e^{iS_I(\frac{\delta}{\delta J})} {\mathbb Z}_0[J]=
\int ({\cal D}\varphi)\, e^{iS(\varphi)+i(J,\,\varphi)},
\\
\label{GenFun-1-2}
&&{\mathbb Z}_0[J]={\cal N}e^{(J,\Delta_F \, J)}=
\int ({\cal D}\varphi)\, e^{iS_0(\varphi)+i(J,\,\varphi)},
\end{eqnarray}
where $\Delta_F$ implies the Feynman propagator;
$S(\varphi)=S_0(\varphi; m)+S_I(\varphi)$ denotes the sum of free and interaction actions
\footnote{For the sake of shortness, we use the notations
$(a,K b)=\int dz_1\, dz_2 a(z_1) K(z_1,z_2)b(z_2)$}.
The stationary phase method applied to ${\mathbb Z}[J]$ gives the following
series (cf.~\cite{Peskin:1995ev, Vasilev:1998})
\begin{eqnarray}
\label{Z-st-ph}
&&{\mathbb Z}[J]= e^{iS(\varphi_c) + i (J,\,\varphi_c)} \int ({\cal D}\eta)
 e^{-\frac{i}{2}(\eta,\, \Box\eta)}
 \text{exp} \Big\{ -i\sum_{n=2}^4 \frac{[\lambda]_n}{n!} \big( 1, \eta^n \big)\Big\}
\nonumber\\
&&
= e^{iS(\varphi_c) + i (J,\,\varphi_c)}\, P_\eta \,\text{exp}\big\{ V(\eta)\big\} \Big|_{\eta=0}
\quad \text{with} \quad P_\eta\equiv \text{exp} \Big\{ \frac{1}{2} (\frac{\delta}{\delta \eta}, \Delta_F \frac{\delta}{\delta \eta} )\Big\}
\end{eqnarray}
where $\eta=\varphi - \varphi_c$ with
$\lim_{J\to 0} \varphi_c(x)\equiv \lim_{J\to 0}  \langle 0| \varphi |0 \rangle^J = \varphi_c=const$.
The classical field configuration, denoted by $\varphi_c(x)$, corresponds to the solution of $\delta S(\varphi)/\delta \varphi(x)=0$.

Notice that this expansion should actually be considered as an asymptotical series
and all inner lines correspond to the scalar {\it massless} propagators.
Besides, the generating function of Eqn.~(\ref{Z-st-ph}) generates the vertices which are
\begin{eqnarray}
\label{ver-1}
   &(a)&\Rightarrow    [\lambda]_2 \eta^2 \equiv \lambda_0^{(a)} \eta^2   \stackrel{\text{def}}{=}  \big( m_0^2+\lambda_0\varphi_c^2 /2\big) \eta^2;
\nonumber\\
   &(b)& \Rightarrow  [\lambda]_3 \eta^3 \equiv \lambda_0^{(b)} \eta^3   \stackrel{\text{def}}{=} \lambda_0\varphi_c \eta^3;
\nonumber\\
   &(c)&\Rightarrow  [\lambda]_4 \eta^4  \stackrel{\text{def}}{=} \lambda_0\eta^4.
\end{eqnarray}
In Eqn.~(\ref{ver-1}), the mass and coupling constant (charge) are bare ones.
It is worth to note that the vertices $(a)$ and $(b)$ should be treated
as effective ones, while $(c)$ corresponds to the standard vertex in the $\varphi^4$-theory under consideration.

The connected generalizing functional ${\mathbb W}[J]$ is related to the effective action $\varGamma[\varphi]$ as
(the Legendre transformations)
\begin{eqnarray}
\label{W-G-def}
\varGamma[\varphi] = {\mathbb W}[J] - i(J,\,\varphi).
\end{eqnarray}
Based on the generating functional, see Eqn.~(\ref{Z-st-ph}), and on the Legendre transform,
see Eqn.~(\ref{W-G-def}), we can readily derive the expression
for the effective action/potential. 
Symbolically, we have
\begin{eqnarray}
\label{Eff-Pon-G-1}
\varGamma[\varphi_c]=S(\varphi_c)
+ \ln\big[ (\text{det}\,\widehat{\Box})^{-1/2}\big] + \,
\Big\{  n\text{-loop connected diagrams} \Big\},
\end{eqnarray}
where the term of $\ln\big[ (\text{det}\,\widehat{\Box})^{-1/2}\big]$,
which corresponds to the one-loop standard diagram contribution only,
does not actually contribute in the massless propagator case.
While, the second term of Eqn.~(\ref{Eff-Pon-G-1}) involves the full set of the connected diagrams
which can be grouped as follows:
{\it (a)} the standard diagrams in $\varphi^4$ with the $[\lambda]^n$-vertices only
\footnote{The standard vacuum diagrams with $[\lambda]^n$-vertices do not depend on $\varphi_c$ and, therefore,
they can be omitted at the moment.};
{\it (b)} the non-standard diagrams of type-$I$ with the $[\lambda^{(a)}]^n$-vertices only;
{\it (c)} the non-standard diagrams of type-$II$ with the $[\lambda^{(b)}]^{2n}$-vertices only;
{\it (d)} the diagrams of type-$III$ with the mixed vertices as
$[\lambda^{(a)}]^{n_1} [\lambda^{(b)}]^{n_2} [\lambda]^{n_3}$.

\begin{figure}[t]
\centerline{\includegraphics[width=0.08\textwidth]{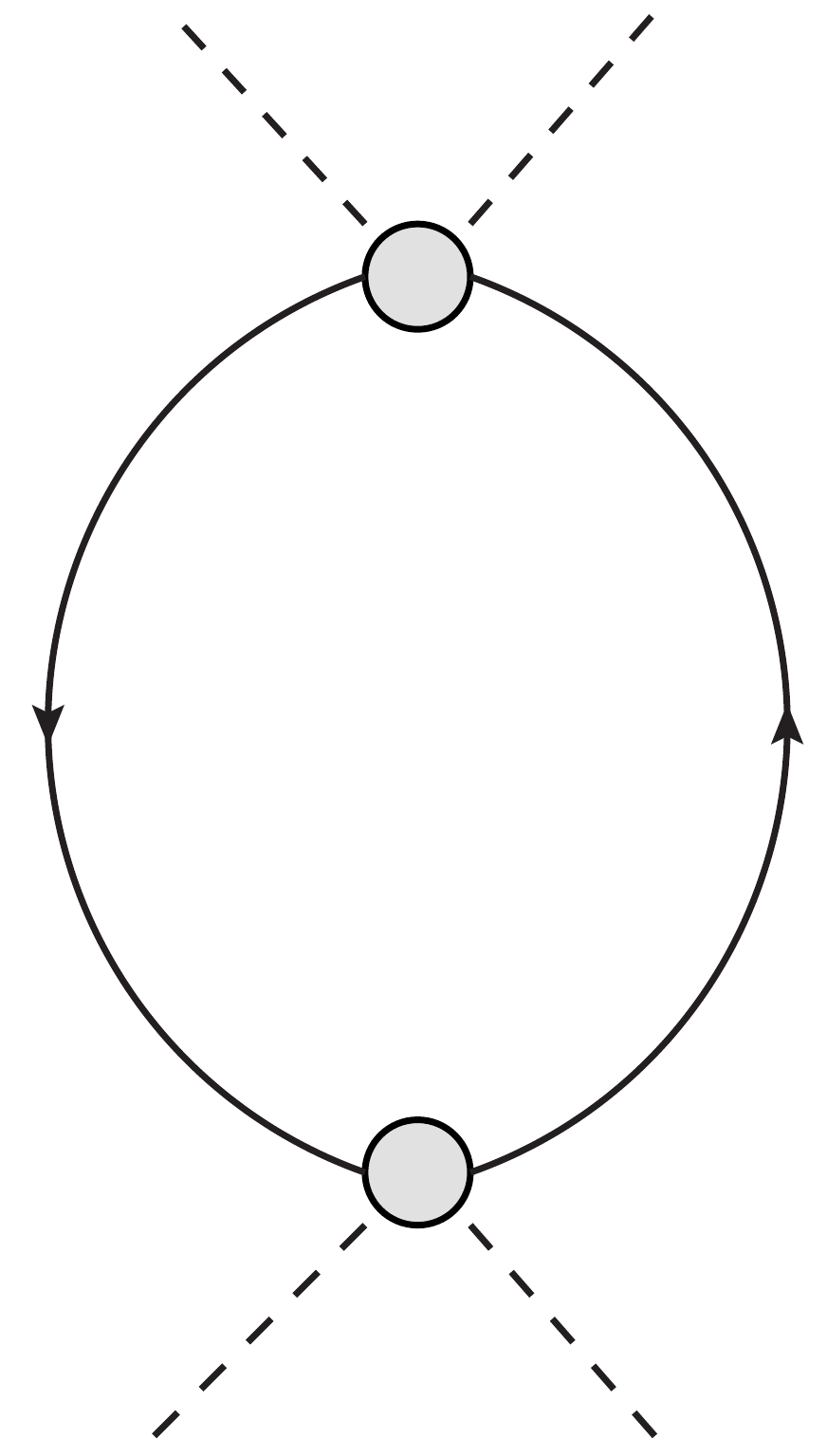}\hspace{0.4cm} \includegraphics[width=0.15\textwidth]{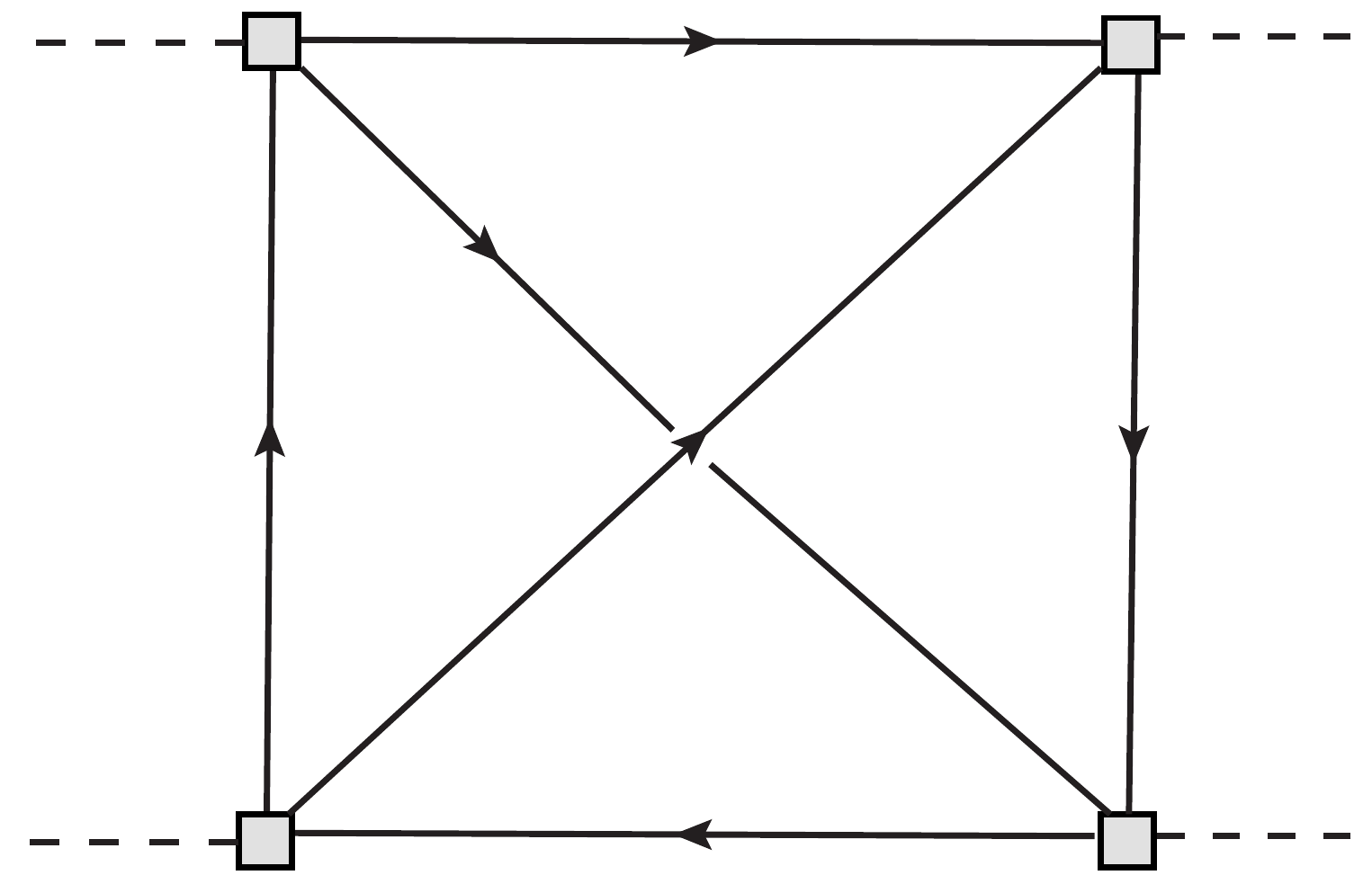}\hspace{0.4cm}
\includegraphics[width=0.15\textwidth]{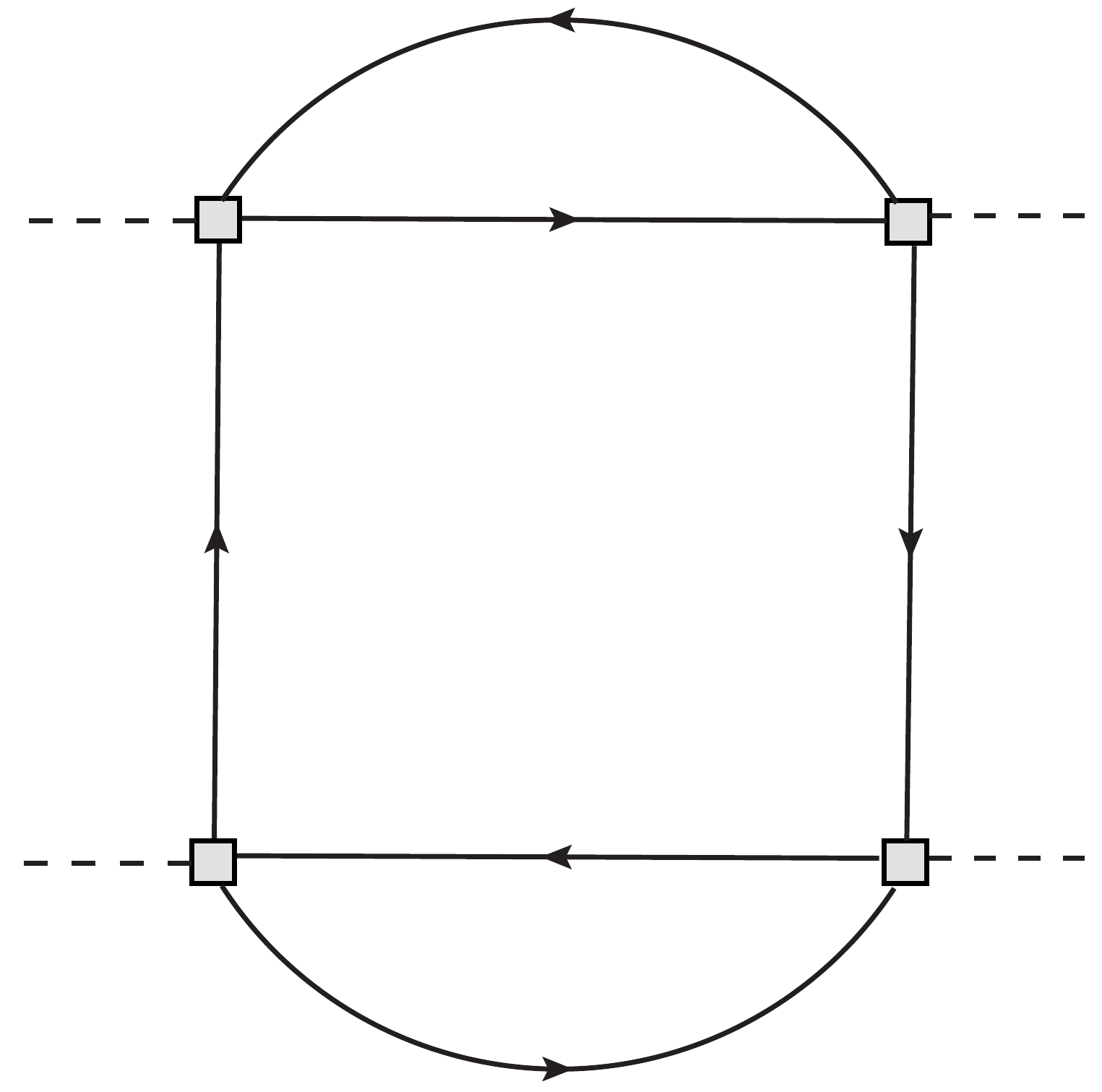}\hspace{0.4cm}
\includegraphics[width=0.15\textwidth]{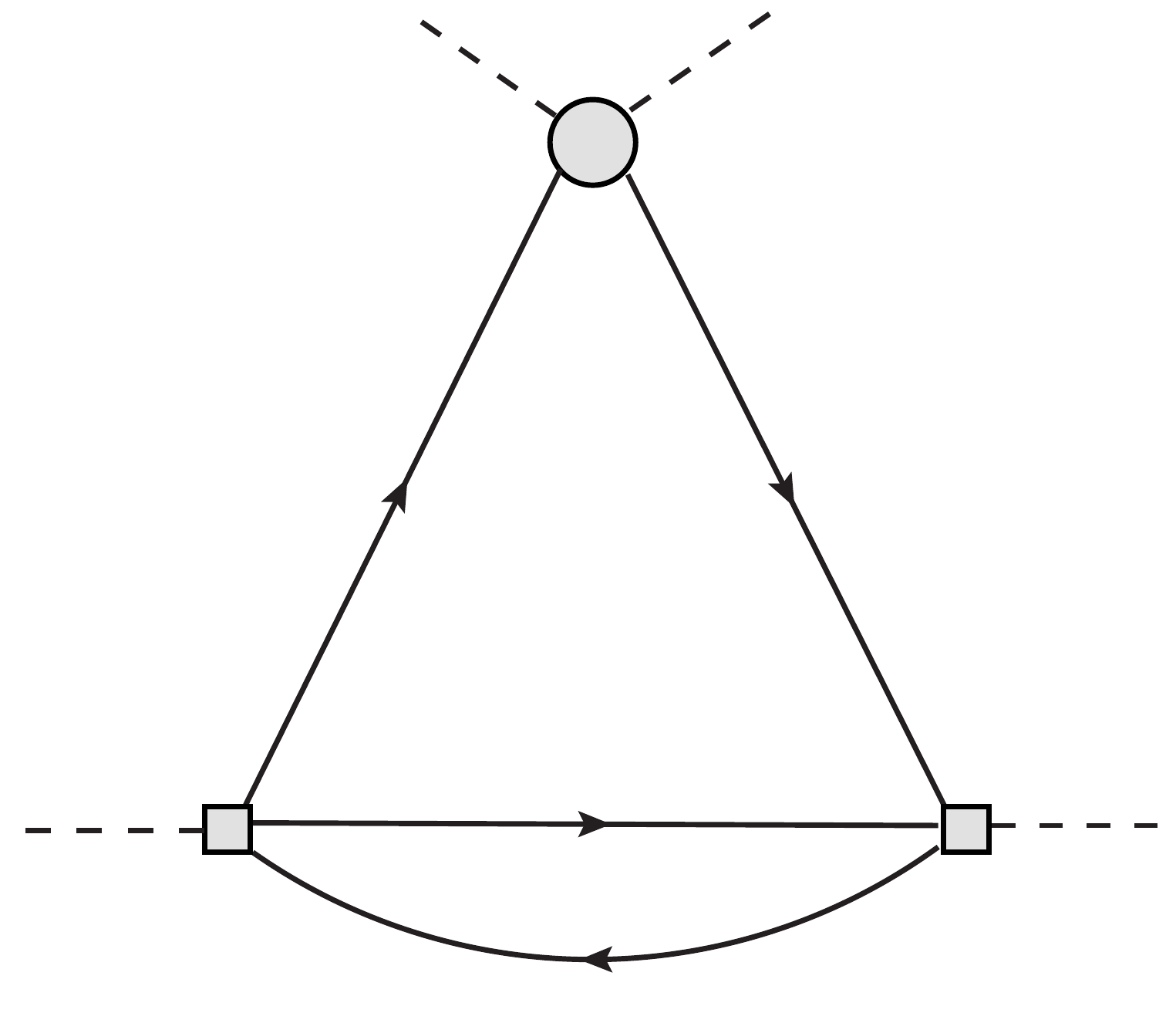} \hspace{0.4cm} \includegraphics[width=0.1\textwidth]{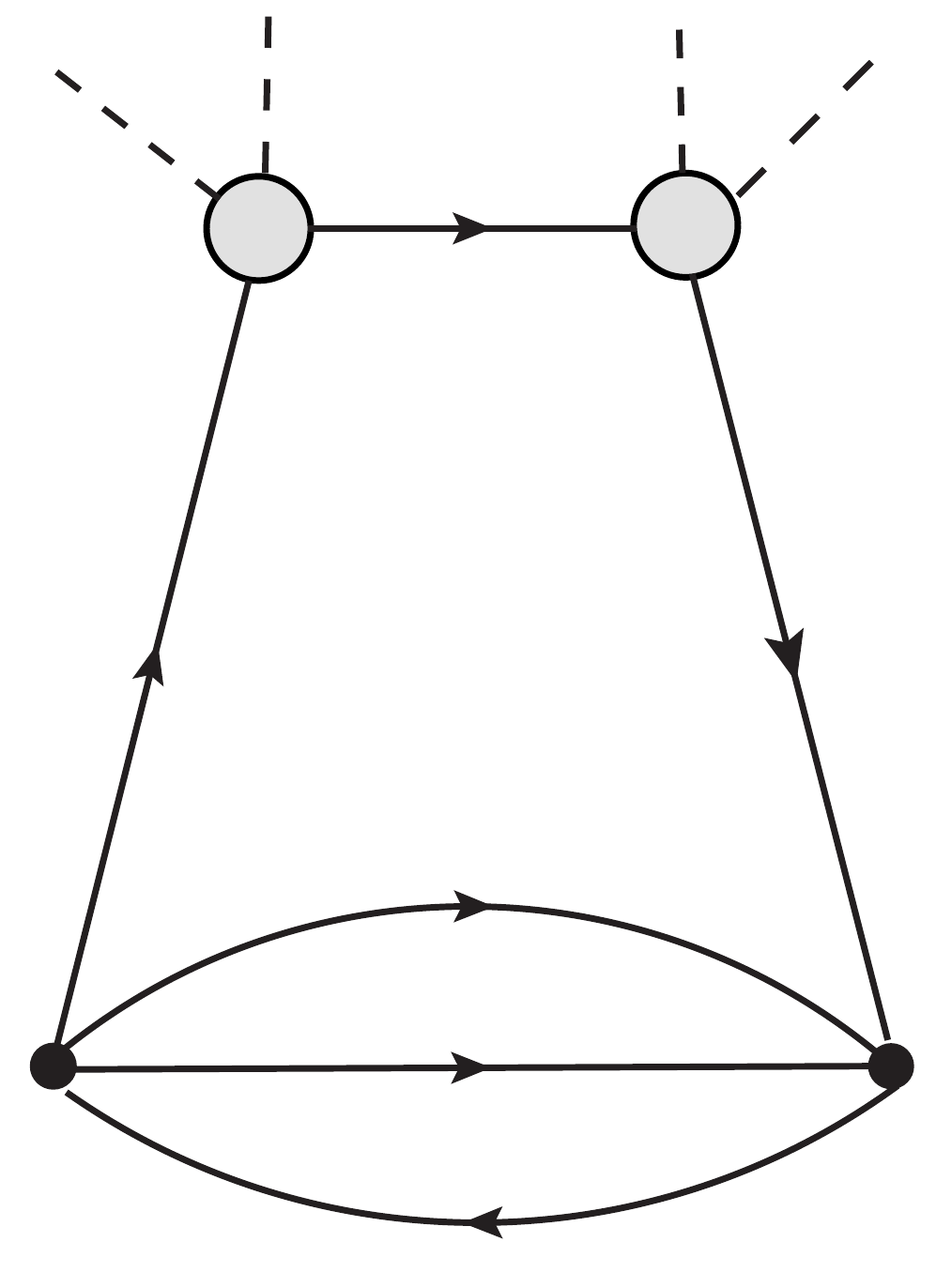}}
\caption{The diagrams contributing up to the order of $[\lambda]^4$.}
\label{Fig-W-1}
\end{figure}

The non-standard diagrams of type-$I$ contribute only to the one-loop approximation.
In this case, we have the only contribution as \cite{Gorishnii:1984te}
(see the first diagram of Fig.~\ref{Fig-W-1})
\begin{eqnarray}
\label{diaI-1}
\varGamma^{(I)}[\varphi_c]=\sum_{n=1}^{\infty} \int (d^D k) \frac{[\lambda_0^{(a)}]^n}{(k^2)^n} =
\sum_{n=1}^{\infty} \frac{[\lambda_0^{(a)}]^n}{\Gamma(n)} \delta\big(n-D/2 \big)=
\frac{ [\lambda_0^{(a)}]^{D/2}}{\Gamma(D/2)} \delta(0).
\end{eqnarray}
Here and in what follows, the singularity of $\delta(0)$ should be treated
as (see, for example, \cite{Anikin:2020dlh, Antosik:1973})
\begin{eqnarray}
\label{delta-sing}
\delta(0)\equiv \lim_{\varepsilon\to 0} \frac{a_{(I)}}{\varepsilon},
\end{eqnarray}
where $a_{(I)}$ is, generally speaking, an arbitrary constant
\footnote{The constants $a_{(i)}$ corresponding to the given diagram involves also the diagram symmetric coefficient.}
and it can be fixed by the pole relations, see below.
Moreover, the pre-delta function can be $\varepsilon$-expanded  \cite{Anikin:2020dlh}.

The representations given by Eqns.~(\ref{diaI-1}) and (\ref{delta-sing}) require some explanations.
First of all, in the vacuum integration series of (\ref{diaI-1}) we focus on the ultraviolet divergency
only, otherwise the vacuum massless integrations are nullified after the infrared divergency has been included.
Then, $\varGamma^{(I)}[\varphi_c]$ receives the only contribution which goes from the following integration \cite{Grozin:2005yg} (here $D=4-2\varepsilon$)
\begin{eqnarray}
\label{delta-sing-2}
&&\varGamma^{(I)}[\varphi_c] =
[\lambda_0^{(a)}]^2
\int_{\text{UV}} \frac{(d^D k) }{(k^2)^2} \equiv
[\lambda_0^{(a)}]^2 \frac{\pi^{D/2}}{\Gamma(D/2)} \int_{\mu^2}^{\infty} d\beta \beta^{D/2-3}=
\nonumber\\
&&
[\lambda_0^{(a)}]^2 \frac{\pi^{2-\varepsilon}}{\Gamma(2-\varepsilon)}
\frac{\mu^{-2\varepsilon}}{\varepsilon}\Big|_{\varepsilon\to 0}=
[\lambda_0^{(a)}]^{2-\varepsilon} \frac{\pi^{2-\varepsilon}}{\Gamma(2-\varepsilon)}\,
\frac{1}{\varepsilon}\Big|_{\varepsilon\to 0},
\end{eqnarray}
where $\beta=|k|^2$ and $\mu^2$ has been chosen to be equal to $\lambda_0^{(a)}$.

On the other hand,
let us calculate the series related to $\varGamma^{(I)}[\varphi_c]$ with the help of the vacuum integration technique \cite{Gorishnii:1984te}.
We obtain that (cf. Eqn.~(\ref{diaI-1}))
\begin{eqnarray}
\label{diaI-1-2}
&&\varGamma^{(I)}[\varphi_c]=
\sum_{n=1}^{\infty} \int (d^D k) \frac{[\lambda_0^{(a)}]^n}{(k^2)^n} =
\frac{[\lambda_0^{(a)}]^{D/2}}{\Gamma(D/2)}\sum_{n=1}^{\infty} \delta\big(n-D/2\big)
\nonumber\\
&&
=\frac{[\lambda_0^{(a)}]^{2-\varepsilon}}{\Gamma(2-\varepsilon)}\sum_{n=1}^{\infty} \delta\big(n-2+\varepsilon\big)=
\frac{[\lambda_0^{(a)}]^{2-\varepsilon}}{\Gamma(2-\varepsilon)}\delta\big(\varepsilon\big)
\end{eqnarray}
Eqn.~(\ref{diaI-1-2}) involves the singular generated function (distribution) $\delta(\varepsilon)$ that is the well-defined
functional on the finite $\phi$-function space with the integration measure $d\mu(\varepsilon)=d\varepsilon \, \phi(\varepsilon)$.
Nonetheless, in many cases it is not convenient, from the technical viewpoint, to introduce the space with the measure $d\mu(\varepsilon)$.
Eqns.~(\ref{diaI-1}) and (\ref{diaI-1-2}) should be equivalent (these equations are merely different representations of the given diagram),
it hints to use the sequential approach \cite{Antosik:1973} to the delta-function and, as consequence, to the treatment of
$\delta(0)$-singularity/uncertintity.
In other words, we may say that $\delta(0)$ is only a symbol of the limit given by $\lim_{\varepsilon\to 0}[1/\varepsilon]$.

So, we can make an inference that after the $\delta(0)$-singularity has been singled out
({\it i.e.} the vacuum integration has been implemented in Eqn.~(\ref{diaI-1})),
the dimension $D$ as an argument of $\Gamma$-function can be extended to
$D=4-2\varepsilon$, see \cite{Anikin:2020dlh, Grozin:2005yg}, giving the non-trivial $\varepsilon$-expansion.
In this paper, it is, however, enough to be restricted by the region of $\varepsilon=0$ in the corresponding pre-delta functions.

The full set of type-$II$ diagrams reduces to the three-loop box-like diagram,
see the second and third diagrams of Fig.~\ref{Fig-W-1}. It reads
\begin{eqnarray}
\label{diaII-1}
\varGamma^{(II)}[\varphi_c]=
\Big\{ G(1,1,1,1,1)+ 2G^2(1,1)\Big\} \, \frac{ [\lambda_0^{(b)}]^{4}}{\Gamma(2)} \delta(3 \varepsilon),
\end{eqnarray}
where
\begin{eqnarray}
\label{G-4-1}
G(1,1,1,1,1)&=&\frac{2}{D-4} \Big\{ -(D-3) G^2(1,1)
\nonumber\\
&+&
\frac{(3D-8)(3D-10)}{D-4} G(1,1) G(1, 2-D/2)\Big\},
\\
\label{G-1-1}
G(N_1, N_2)&=&\frac{\Gamma(N_1+N_2-D/2) \, \Gamma(D/2-N_1)\Gamma(D/2-N_2)}{\Gamma(N_1)\Gamma(N_2)\, \Gamma(D-N_1-N_2)}.
\end{eqnarray}
Indeed, the general structure of this sum can be symbolically presented as
\begin{eqnarray}
\label{diaII-1-2}
\varGamma^{(II)}[\varphi_c]\sim \sum_{n=1}^{\infty}[\lambda_0^{(b)}]^{2n}
\, \delta\big(3 n - (n+1)D/2 \big).
\end{eqnarray}
From Eqn.~(\ref{diaII-1-2}), one can immediately conclude that the only contribution originates from the case of $n=2$,
{\it i.e.} $[\lambda_0^{(b)}]^4$, giving the delta function as
$\delta(6-3D/2) \sim \delta(\varepsilon)$. The other contributions correspond to the delta function where 
the argument ensures the zero contributions.
However, this type of diagrams can be omitted at this moment due to the highest singularity that behaves as $1/\varepsilon^3$.

The mixed diagrams of type-$III$ can be aggregated into two classes.
The first class $A$ of diagrams with $n_1=n,\, n_2=2, \, n_3=0$ leads to
two-loop contributions (see the fourth diagram of Fig.~\ref{Fig-W-1}) 
which are given by
\begin{eqnarray}
\label{diaIII-1}
\varGamma^{(III)}_{A}[\varphi_c]&=&[\lambda_0^{(b)}]^{2}\, \sum_{n=1}^{\infty} \int (d^D k) \frac{[\lambda_0^{(a)}]^n}{(k^2)^{n+1}}
\int  \frac{(d^D \ell)}{\ell^2 (\ell -k)^2}
\sim
[\lambda_0^{(b)}]^{2} \sum_{n=1}^{\infty}[\lambda_0^{(a)}]^n \delta\big(n+3 -D \big)
\nonumber\\
&=&
[\lambda_0^{(b)}]^{2}\, G(1,1) \frac{[\lambda_0^{(a)}]}{\Gamma(2)} \delta(0)
\end{eqnarray}
The second class $B$ of diagrams with $n_1=n,\, n_2=0, \, n_3=2$ can be presented in the form of
three-loop integration 
(see the fifth diagram of Fig.~\ref{Fig-W-1}) as
\begin{eqnarray}
\label{Diam-1}
\varGamma^{(III)}_{B}[\varphi_c]&=&
[\lambda_0]^2 \,
\sum_{n=1}^{\infty} \int (d^D k)\frac{[\lambda_0^{(a)}]^n}{(k^2)^{n+1}} \, \int \frac{(d^D \ell)}{\ell^2}
\int \frac{(d^D p)}{p^2 (k+p-\ell)^2}
\nonumber\\
&\sim& [\lambda_0]^2 \, \sum_{n=1}^{\infty} [\lambda_0^{(a)}]^n \delta\big(n+4 -3D/2 \big)
\nonumber\\
&=& [\lambda_0]^2 \,G(1,1)\, G(1, 2-D/2)\,
\frac{[\lambda_0^{(a)}]^2}{\Gamma(2)}\, \delta(0).
\end{eqnarray}

Thus, to the order of $[\lambda]^4$, the connected diagrams in Eqn.~(\ref{Eff-Pon-G-1}) contribute as
 \begin{eqnarray}
\label{Eff-Pon-G-1-2}
\varGamma_c[\varphi_c]=\varGamma^{(I)}[\varphi_c]+
\varGamma^{(II)}[\varphi_c]+\varGamma^{(III)}_{A}[\varphi_c]+\varGamma^{(III)}_{B}[\varphi_c],
\end{eqnarray}
where $\varGamma^{(i)}[\varphi_c]$ with $i=\{ (I); (II); (III), A \}$ are uniquely fixed in a sense that they
only contribute to the definite order of $[\lambda]^k$ ($k=2,4,3$ respectively).
In contrast, $\varGamma^{(III)}_{B}[\varphi_c]$ can involve the higher order of $[\lambda]^{k}$ with $k\ge 4$.

At $\ell$-loop accuracy, every of connected diagrams
contains both the singular and finite parts. As usual, the singular parts should be eliminated by the corresponding counterterms
within the certain renormalization procedure resulting in the appearance of the dimensional parameter (scale) $\mu$
\footnote{We remind that $\mu$ is related to some subtraction point.}.
Re-expressing via the renormalized and dimensionless charge $\lambda$ (within the dimensional regularization),
we thus have the following effective action/potential
\begin{eqnarray}
\label{Delta-G}
\varGamma[\varphi_c]&=&
\varGamma_2(0)\,\varphi_c^2(x) + \varGamma_4(0)\, \varphi_c^4(x) + ....
\nonumber\\
&=&
 \frac{m_0^2}{2} \Big( 1+  \Delta Z_m(\lambda) \Big)^{-1} \varphi_c^2 +
 \mu^{2\varepsilon} \frac{\lambda_0}{4!}  \Big( 1+  \Delta Z_\lambda(\lambda) \Big)^{-1}\varphi_c^4  + \{\text{finite terms}\},
\end{eqnarray}
where $\varGamma_{2,\, 4}(0)$ imply the $1$PI Green (vertex) functions and
\begin{eqnarray}
\label{DeltaZ}
\Delta Z_{m; \lambda}(\lambda) = \sum_{n=1} \frac{C_n^{\{m; \lambda\}}(\lambda)}{\varepsilon^n}=
\sum_{n=1}\frac{1}{\varepsilon^n} \sum_{k=1} C_{n k}^{\{m; \lambda\}} \lambda^k
\end{eqnarray}
with (see also below in Sec.~\ref{SubSec-PR})
\begin{eqnarray}
\label{C1m}
&&C_1^{\{m\}}(\lambda)= \lambda \Big( c_1^{[m; (I)]} + \lambda c_1^{[m; (III), A]} + \lambda^2 c_1^{[m; (III), B]} \Big)+  o(\lambda^4),
\\
\label{C2m}
&&C_2^{\{m\}}(\lambda)= \lambda^2 \Big[ c_2^{[m; (III), A]} + \lambda c_2^{[m; (III), B]} \Big] + o(\lambda^4),
\\
\label{C1lam}
&&C_1^{\{\lambda\}}(\lambda)= \lambda \Big(  c_1^{[\lambda; (I)]} + \lambda c_1^{[\lambda; (III), A]} +
\lambda^2  c_1^{[\lambda; (III), B]} \Big)
+ o(\lambda^4),
\\
\label{C2lam}
&&C_2^{\{\lambda\}}(\lambda)= \lambda^2 \Big( c_2^{[\lambda; (III), A]} + \lambda c_2^{[\lambda; (III), B]} \Big) + o(\lambda^4).
\end{eqnarray}

As usual, the evolution of finite effective action/potential with respect to the different scale choice is governed by the
anomalous dimension (or Hamiltonian which has, generally speaking, a form of the integral operator).
In its turn, the anomalous dimensions are determined through the coefficients $C_1(\lambda)$, see Eqns.~(\ref{C1m}) and (\ref{C1lam}),
in the pole relations.
In the considered accuracy ( we remind that we deal with corrections up to $[\lambda]^4$-order)
the contributions of all vacuum diagrams can be derived relatively easy.
As a consequence, we are able to obtain the anomalous dimension needed for the evolution rather fast.
However, if we need the higher order of accuracy (even an arbitrary order of accuracy),
the calculations of anomalous dimensions related to some of diagrams might to be tricky.
In this connection, we have found that the contribution of diagram $(III)$ of the second class $B$ to
the anomalous dimension can be computed almost algebraic based on the known anomalous dimension of the corresponding
non-local operator Green function $G^{(2)}_{\cal O}$. This can be done with the help of $V_{z,x}$-operation the description of which is presented
in the next section.
%
%

\section{Vacuum $V_{z,x}$-operation: transformation of Green functions into vacuum integrations}
\label{Sec-Vop-1}

All vacuum integrations can be performed by direct calculations
\cite{Anikin:2020dlh, Gorishnii:1984te}.
However, in the case of $n_1=2$ (or, in other words, if the delta-function, appearing after the vacuum integration,
separates out the only term with $n_1=2$ in the full sum),
the method that is based on the manifestation of conformal symmetry can be extremely useful \cite{Braun:2013tva, Anikin2023}.
Indeed, in \cite{Braun:2013tva} it has been demonstrated that the certain constraints, thanks for
the conformal symmetry, has been encoded in terms of the
generators of the collinear $SL(2)$ subgroup (the algebra of which is isomorphic to $su(2,2)$).
At that, two generators,
denoted as $S_{-}$ and $S_{0}$, can be defined at all loops with the help of the
evolution kernel. At the same time, the special conformal
generator $S_{+}$ involves the nontrivial corrections
and can be calculated order by order in perturbation
theory. If the generator $S_+$ is known at the
order of $(\ell -1)$ loop, the corresponding evolution kernel in
the physical dimension can be fixed to the $\ell$-loop accuracy (up to the terms
which are invariant regarding the tree-level generators).

In \cite{Anikin2023}, it is shown that one can adopt the algebraic approach \cite{Braun:2013tva} to the multi-loop vacuum integrations.
To this goal, the special $V_{z,x}$-operation has  been introduced. We now dwell on
the demonstration of the $V_{z,x}$-operation that transforms the given Green function in
the vacuum integrations. Notice that our $V_{z,x}$-operation can be referred as the inverse operation presented in \cite{Gorishnii:1984te}.

Let us consider the second class $B$ of mixed diagram $(III)$ that contributes as presented by Eqn.~(\ref{Diam-1}).
As mentioned, only this type of diagrams can include the higher order of $[\lambda]$.
The contributions of the other three diagrams $(I)$, $(II)$ and the class $A$ of $(III)$,
see Eqns.~(\ref{diaI-1}), (\ref{diaII-1}) and (\ref{diaIII-1}), are fixed and they stay unchanged depending on the definite order of $[\lambda]$.

We now define the $V_{z,x}$-procedure as
\begin{eqnarray}
\label{Diam-1-2}
\overline{\varGamma}^{(2)}_{(III),\, B}[\varphi_c]= \frac{1}{C^{(2)}(D)}\,V_{z,x} \Big\{
G^{(2)}_{\cal{O}}(x_1, x_2 ; z_1, z_2)\Big\},
\end{eqnarray}
where
\begin{eqnarray}
\label{BarGamma}
&&[\lambda^{(a)}]^{3D/2-4}\, \overline{\varGamma}^{(2)}_{(III),\, B}[\varphi_c]  = \varGamma^{(2)}_{(III),\, B}[\varphi_c],
\\
\label{CD}
&&C^{(2)}(D)=(D/3-1)\, \frac{\Gamma(D/2)}{\Gamma(6-D)} \,
\frac{\prod\limits_{\kappa_1=4}^{6} \big( 3D/2- \kappa_1\big)}
{\prod\limits_{\kappa_2=2}^{4} \big( D/2- \kappa_2\big)}
\end{eqnarray}
and
\begin{eqnarray}
\label{Diam-1-2-3}
&&V_{z,x} \Big\{
G^{(2)}_{\cal{O}}(x_1, x_2 ; z_1, z_2)\Big\}
\equiv V_{z} \Big\{  V_{x} \Big[
G^{(2)}_{\cal{O}}(x_1, x_2 ; z_1, z_2) \Big]\Big\}
 \stackrel{\text{def}}{=}
\\
&&
\int d^D z_1\, d^D z_2 \Delta_F(z_1-z_2) \Big[
\int d^D x_1 d^D x_2 \delta(x_1-x_2) \widehat{\Box}_{x_2} \,
\Big\{ G^{(2)}_{\cal{O}}(x_1, x_2 ; z_1, z_2)\Big\} \Big].
\nonumber
\end{eqnarray}
In Eqn.~(\ref{Diam-1-2-3}), to $[\lambda]^2$-order of $\varphi^4$-interaction,
the Green function $G^{(2)}_{\cal{O}}(x_1, x_2 ; z_1, z_2)$ reads
\begin{eqnarray}
\label{GF-1}
G^{(2)}_{\cal{O}}(x_1, x_2 ; z_1, z_2) =
\langle 0| T \eta(x_1) \eta(x_2)\, {\cal O}(z_1, z_2)
\Big[ [\lambda] \int d^D y \eta^4(y)\Big]^2 |0\rangle,
\end{eqnarray}
with the inserted non-local operator ${\cal O}(z_1, z_2)=\eta(z_1)\eta(z_2)$.
The coupling constant denoted as $[\lambda]$ is absorbing the combinatory factor which is now irrelevant for our consideration.

We are now in a position to proof the statement expressed by Eqn.~(\ref{Diam-1-2}).
We begin with the momentum representation ($p$-space) where
the Green function $G^{(2)}_{\cal{O}}(x_1, x_2 ; z_1, z_2)$ represented by the diagram depicted in Fig.~\ref{Fig-W-3}
and it takes a form of
\begin{eqnarray}
\label{GF-O-1}
&&G^{(2)}_{\cal{O}}(x_1, x_2 ; z_1, z_2) =
\nonumber\\
&&
\int (d^D k_1)\, (d^D k_2) e^{-ik_1 x_1 +ik_2 x_2}
S(k_1) S(k_2)
 \int (d^D k_3)\, (d^D k_4) e^{+ik_3 z_1 - ik_4 z_2}
S(k_3) S(k_4)
\nonumber\\
&&\times
 \int (d^D p)\, S(p) S(p+k_1-k_3) \, \delta^{(D)}\big( k_1+k_4 - k_2- k_3\big).
\end{eqnarray}
After some algebra, we obtain that
\begin{eqnarray}
\label{GF-O-2}
G^{(2)}_{\cal{O}}(x_1, x_2 ; z_1, z_2) &=&
G(1,1) \frac{\Gamma(4-D)}{\Gamma(2-D/2)}
\int_0^1 d\mu(\alpha) \int (d^D k_1)\, (d^D k_2) e^{-ik_1 (x_1-z^{\alpha_1}_{12}) +ik_2 (x_2-z^{\alpha_2}_{21})}
\nonumber\\
&&\times
S(k_1) S(k_2)
\big[
\mathbb{B}(k_1,k_2, \alpha)
 \big]^{D-4},
\end{eqnarray}
where
$z^{\alpha}_{12}=\overline{\alpha}z_1 + \alpha z_2$, and
\begin{eqnarray}
\label{B-func}
\mathbb{B}(k_1,k_2, \alpha)=
\alpha_2 \overline{\alpha}_2 (k_2-k_1)^2 + \alpha_3 \overline{\alpha}_3 k_1^2
+ \alpha_2 \alpha_3 (k_2-k_1)k_1.
\end{eqnarray}
In Eqn.~(\ref{GF-O-2}),
the integration measure in $\alpha$-space is given by
\begin{eqnarray}
\label{dmu}
d\mu(\alpha)=d\alpha_1\, d\alpha_2\, d\alpha_3 \,\delta\Big(1-\sum_{i=1}^{3} \alpha_i \Big) \,\alpha_3^{1-D/2}.
\end{eqnarray}
Then, we apply $V_x$-operator on the Green function that results in
\begin{eqnarray}
\label{Vx}
&&V_{x} \Big[
G^{(2)}_{\cal{O}}(x_1, x_2 ; z_1, z_2) \Big]\equiv
\int d^D x_1 d^D x_2 \delta(x_1-x_2) \widehat{\Box}_{x_2} \,
\Big[ G^{(2)}_{\cal{O}}(x_1, x_2 ; z_1, z_2) \Big]=
\nonumber\\
&&G(1,1) \frac{\Gamma(4-D)}{\Gamma(2-D/2)}
\int_0^1 \frac{d\mu(\alpha)}{\big[ \alpha_3\overline{\alpha}_3\big]^{4-D}} \int (d^D k_1)\,
\frac{e^{+i\alpha_3 k_1 (z_1-z_2)}}{[ k_1^2 ]^{5-D}}.
\end{eqnarray}
Ultimately, the action of $V_z$-operator leads to the following representation:
\begin{eqnarray}
\label{Vzx}
&&V_z\Big\{ V_{x} \Big[
G^{(2)}_{\cal{O}}(x_1, x_2 ; z_1, z_2) \Big]\Big\} \equiv
 \int d^D z_1\, d^D z_2 \Delta_F(z_1-z_2) \Big\{ V_{x} \Big[
G^{(2)}_{\cal{O}}(x_1, x_2 ; z_1, z_2) \Big] \Big\}
\nonumber\\
&&= G(1,1) \frac{\Gamma(4-D)}{\Gamma(2-D/2)}
\frac{\Gamma(D/2-4) \Gamma(D-2)}{\Gamma(3D/2-6)} \frac{\delta(6-3D/2)}{\Gamma(6-D)}.
\end{eqnarray}
%
%
\begin{figure}[t]
\centerline{\includegraphics[width=0.25\textwidth]{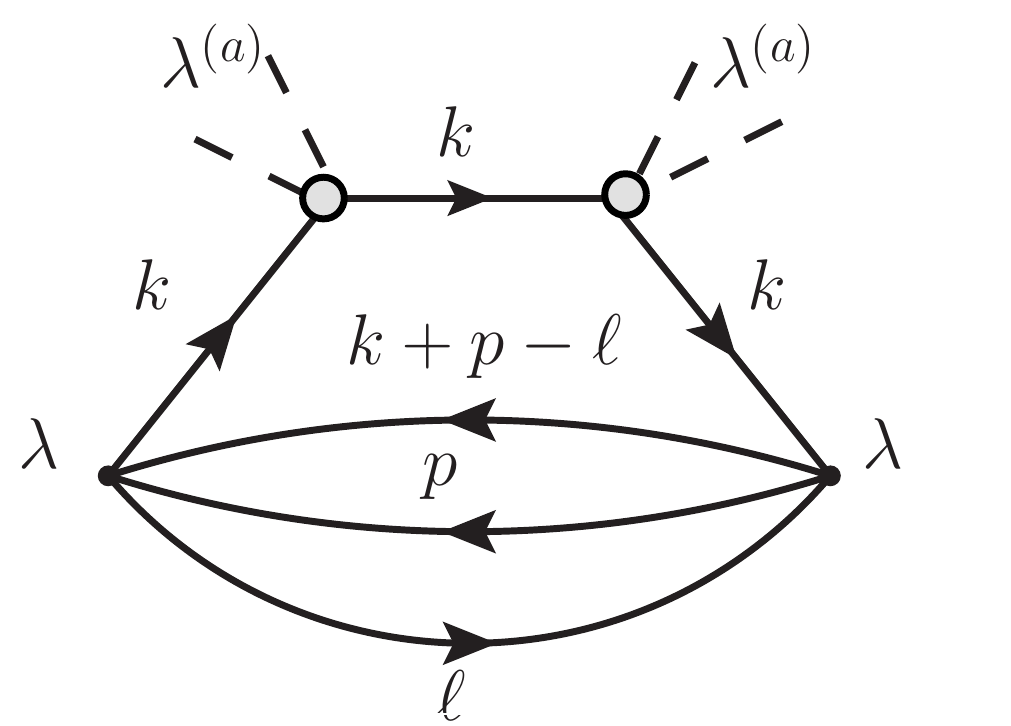}}
\caption{The diagram with $[\lambda]^2 [\lambda^{(a)}]^2$-vertices.}
\label{Fig-W-2}
\end{figure}

\noindent
Multiplying Eqn.~(\ref{Vzx}) by $[\lambda^{(a)}]^{3D/2-4}$ and, then, inserting the result
into the {\it r.h.s.} of Eqn.~(\ref{Diam-1-2}), we obtain the corresponding effective potential
which is represented by Eqn.~(\ref{Diam-1}) .
In Eqn.~(\ref{Vzx}), the dimension $D\in \mathbb{R}$, {\it i.e.}  $D=4-2\varepsilon$,
but the delta-function $\delta(6-3D/2)=\delta(3\varepsilon)$ extracts only the highest singularity in
$\varepsilon$-expansion of the pre-delta $\Gamma$-function combination.

As the next step, including the prefactor $[\lambda^{(a)}]^{3D/2-4}$
we have to make an expansion over $\varepsilon$. As a result, we derive the following expression
\begin{eqnarray}
\label{Diam-2-2}
\varGamma_{(III),\, B}^{(2)}[\varphi_c]&=&
[\lambda^{(a)}]^{3D/2-4}\, \overline{\varGamma}_{(III),\, B}^{(2)}[\varphi_c]= \frac{[\lambda^{(a)}]^{3D/2-4}\,}{C^{(2)}(D)}\,V_{z,x} \Big\{
G^{(2)}_{\cal{O}}(x_1, x_2 ; z_1, z_2)\Big\} \Big|_{\varepsilon\to 0}
\nonumber\\
&\stackrel{\mathcal{R}^\prime}{=}& [\lambda^{(a)}]^2 \Big\{
\varGamma_{(III),\, B}^{(2)\, {\rm sing.}}[\varphi_c] + \varGamma_{(III),\, B}^{(2)\, {\rm fin.}}[\varphi_c] \Big\},
\end{eqnarray}
where $\mathcal{R}^\prime$ implies that the $\mathcal{R}^\prime$-operation has been used, and
\begin{eqnarray}
\label{G-sing}
&&\varGamma_{(III),\, B}^{(2)\, {\rm sing.}}[\varphi_c] = \frac{c^{(III),\, B}_2}{\varepsilon^2} +
\frac{c^{(III),\, B}_1}{\varepsilon},
\\
\label{G-fin}
&&\varGamma_{(III),\, B}^{(2)\, {\rm fin.}}[\varphi_c] = \tilde c^{(III),\, B}_0 +
\tilde c^{(III),\, B}_1\ln\frac{[\lambda^{(a)}]}{\mu^2} + \tilde c^{(III),\, B}_2\ln^2\frac{[\lambda^{(a)}]}{\mu^2},
\\
&&\ln\frac{[\lambda^{(a)}]}{\mu^2} = \ln\frac{m^2}{\mu^2} + \sum_{n=1}^{\infty} \frac{[\lambda]^n}{n}
\Big( \frac{\varphi_c^2}{m^2} \Big)^n.
\end{eqnarray}
We remind that $\mathcal{R}^\prime$ defines the $R$-operation without the last subtraction \cite{Vasilev:1998}.

In Eqn.~(\ref{Diam-2-2}), the local singular part of effective potential should be cancelled by the introduction of corresponding counterterms ($Z$-factor in RG-method). While, the non-local singular terms related to the $\varepsilon$-expansion of $[\lambda^{(a)}]^{\ell\, \varepsilon}$
have to be eliminated by the $\mathcal{R}^\prime$-procedure.
Notice that Eqns.~(\ref{Diam-1-2}) and (\ref{Diam-2-2}) can be generated up to any order of $\lambda$ (multi-loop accuracy).

\section{Vacuum $V_{z,x}$-operation and  anomalous dimensions}
\label{Sec-AD}

The advantage of representation~(\ref{Diam-1-2}) is that assuming the anomalous dimension (evolution kernel) of
the non-local operator Green functions is known at $\ell$-loop accuracy,
one can replace the direct calculations of evolution kernels associated with the corresponding vacuum diagrams
at $(\ell+1)$ loop accuracy by rather a simple (mostly algebraic) $V_{z,x}$-operation.
%
%
\begin{figure}[t]
\centerline{\includegraphics[width=0.3\textwidth]{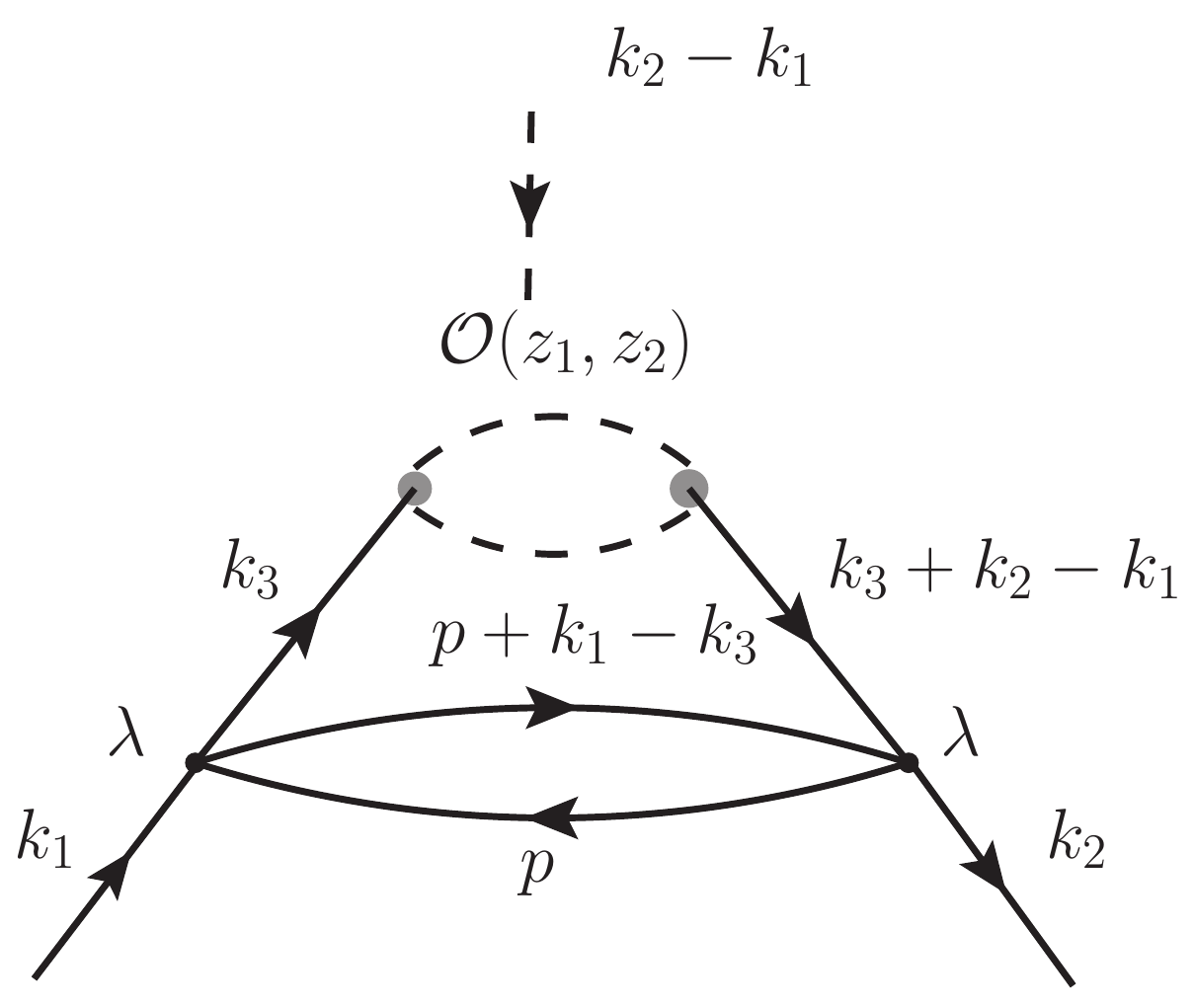}}
\caption{The diagram of $G^{(2)}_{\cal{O}}(x_1, x_2 ; z_1, z_2)$ at the order of $[\lambda]^2$.}
\label{Fig-W-3}
\end{figure}

Indeed, let us return again to the non-local operator Green function at the second order of $[\lambda]$,
see Eqn.~(\ref{GF-O-2}). We now extract the anomalous dimension given by the coefficient $C^{(G_{\cal O})}_1$ if
the corresponding $Z^{(G_{\cal O})}$-factor takes the form of
\begin{eqnarray}
\label{Z-fac}
Z^{(G_{\cal O})}=1+ \sum_{n=1}^\infty \frac{C^{(G_{\cal O})}_n(\lambda)}{\varepsilon^n}.
\end{eqnarray}
To our aim, it is enough to make a replacement as $\mathbb{B}(k_1, k_2, \alpha)\to 1$ because the $\varepsilon$-expansion of
$\mathbb{B}^{\varepsilon}$ does not affect our extraction procedure.  So,
after $\varepsilon$-expansion of $\Gamma$-combination,
we get that
\begin{eqnarray}
\label{H-1}
G^{(2), 1/\varepsilon}_{\cal{O}}(x_1, x_2 ; z_1, z_2) =
\frac{1}{2\varepsilon} \,
\int_0^1 d\mu(\alpha) \Delta_F(x_1-z^{\alpha_1}_{12}) \Delta_F(z^{\alpha_2}_{21}-x_2).
\end{eqnarray}
We stress that, at the given accuracy, this $1/\varepsilon$-term is a highest singular term because
of the non-locality of considered operator $\mathcal{O}$.

The {\it r.h.s} of Eqn.~(\ref{H-1}) (modulo the coupling constant prefactor which is omitted)
coincides with the corresponding expression for the evolution kernel in $\varphi^4$-theory \cite{Braun:2013tva}
provided by the suitable replacements.  Indeed, we have the following relation
\begin{eqnarray}
\label{H-2}
G^{(2), 1/\varepsilon}_{\cal{O}}(x_1, x_2 ; z_1, z_2)
\Big |^{ \Delta_F(x_1-z^{\alpha_1}_{12}) \to \varphi(z^{\alpha_1}_{12})}_{\Delta_F(z^{\alpha_2}_{21}-x_2)\to \varphi(z^{\alpha_2}_{21})} =
\big[ \mathbb{H}_{12} \, {\cal O}\big] (z_1, z_2).
\end{eqnarray}
which represents the needed matching.
In Eqn.~(\ref{H-2}), $\big[ \mathbb{H}_{12} \, {\cal O}\big]$ implies the two-particle evolution (integral) kernel defined in \cite{Braun:2013tva}.

It is interesting to
notice that since the $\mathbb{B}(k_1, k_2, \alpha)$ accumulates the full information on the loop structure of the given Green function,
the same result for $1/\varepsilon$-term, as in Eqn.~(\ref{H-1}), can be derived in $p$-space working directly with the local operator case and with the sub-divergency subtraction.
To show that, we arrange the so-called momentum flux in the diagram of Fig.~\ref{Fig-W-3} with the help of $k_1=0$ and the local limit, $z_1=z_2$.
As a result of this, the diagram takes a propagator-like type and the loop integration is given by
\begin{eqnarray}
\label{K}
&&{\cal K}^{\text{loc.}}(k_2)= \int (d^D k_3) \frac{1}{k_3^2 (k_3+k_2)^2} \Big\{
\int (d^D p) \frac{1}{p^2 (p-k_3)^2} - \frac{1}{\varepsilon}
\Big\} =
\nonumber\\
&&
G(1,1) G(1, 3-D/2) \, \big[k_2^2\big]^{D-4} - \frac{1}{\varepsilon} G(1,1) \, \big[k_2^2\big]^{D/2-2} \approx
\frac{1}{\varepsilon} \Big( \frac{5}{2} - 2 \Big) + ...=
\frac{1}{2\varepsilon} + ...,
\end{eqnarray}
where the ellipses imply the other possible terms in $\varepsilon$-expansion.

Further, applying $V_{z,x}$-operation to $G^{(2)}_{\cal{O}}(x_1, x_2 ; z_1, z_2) $, we obtain that
\begin{eqnarray}
\label{H-3}
&&V_{z,x} \Big\{ G^{(2), 1/\varepsilon}_{\cal{O}}(x_1, x_2 ; z_1, z_2) \Big\} =
\nonumber\\
&&
\frac{1}{2\varepsilon} \int d\mu(\alpha)
\int (d^D z_1) (d^D z_2)  \Delta_F(z_1 - z_2) \Delta_F\big(\alpha_3(z_2-z_1)\big)=
\nonumber\\
&&\frac{1}{4 \varepsilon} \delta^{(D)}(0)\, \delta(\varepsilon)
\end{eqnarray}
where it has been used that
\begin{eqnarray}
\label{v-int-1}
\int (d^D z_1) (d^D z_2)  \Delta_F(z_1 - z_2) \Delta_F\big(\alpha_3(z_2-z_1)\big) =
\delta^{(D)}(0) \frac{1}{\alpha_3^2} \delta(2-D/2).
\end{eqnarray}
The prefactor $\delta^{(D)}(0)$ gives finally the space-time volume $V\times T$ which connects the effective action with the
effective potential.

In Eqn.~(\ref{H-3}), the vacuum integration (\ref{v-int-1}), {\it i.e.} integrations over $z_1$ and $z_2$, leads to $\delta(2-D/2)$ that behaves as $1/\varepsilon$ within the sequential approach if $D=4-2\varepsilon$
(see for example \cite{Anikin:2020dlh, Antosik:1973}).
Therefore, $V_{z,x}$-operation results in the coefficient $c_2$ for the effective potential, see Eqn.~(\ref{G-sing}).

As a last step, in order to obtain the
anomalous dimension (the coefficient $c_1$ of Eqn.~(\ref{G-sing})) for the effective potential $\varGamma^{(2)}[\varphi_c]$,
we have to use the corresponding pole relations, {\it i.e.}
\begin{eqnarray}
\label{PR-1}
c_1={\rm P}_{\varGamma}(c_2), \quad c_2= V_{z,x} \Big\{ \big[ \mathbb{H}_{12} {\cal O}\big](z_1,z_2) \Big\},
\end{eqnarray}
where the corresponding pole relations for the effective potential denoted as ${\rm P}_{\varGamma}$.

\subsection{The pole relations}
\label{SubSec-PR}

To conclude this section, we study the important consequences of the pole relations that not only relate the different coefficient $c_i$, but
they can fix the arbitrary constants $a_{(i)}$, see Eqn.~(\ref{delta-sing}).

We begin with the schematic derivation of the pole relations.
The pole relations for $\varGamma[\varphi_c]$ are stemmed from
the $\mu\partial_\mu$-differentiation of the effective potential $Z$-factors, $Z_m$ and $Z_\lambda$, defined as
\begin{eqnarray}
\label{EffP-Z-1-2}
\varGamma_0[\varphi_c] = Z^{\varGamma[\varphi_c]} \varGamma[\varphi_c],
\quad
Z^{\varGamma[\varphi_c]} = 1+ \sum_{n=1}^\infty \frac{C_n([\lambda])}{\varepsilon^n}.
\end{eqnarray}
Having calculated $\mu\partial_\mu$-derivative of $Z$-factor, see Eqn.~(\ref{EffP-Z-1-2}), we obtain that
\begin{eqnarray}
\label{EffP-P-1}
&&\Big\{1+ \sum_{n=1}^\infty \frac{C_n([\lambda])}{\varepsilon^n} \Big\} \gamma_{\varGamma[\varphi_c]} =
\beta_\lambda ([\lambda])\partial_\lambda \sum_{n=1}^\infty \frac{C_n([\lambda])}{\varepsilon^n}
\nonumber\\
&&\text{with} \,\,\, 
\beta_\lambda ([\lambda])=  \mu\partial_\mu [\lambda],
\,\,\,\gamma_{\varGamma[\varphi_c]}\equiv  \mu\partial_\mu  \ln Z^{\varGamma[\varphi_c]}
\end{eqnarray}
and, as a consequence, we have the following pole relations
\begin{eqnarray}
\label{EffP-P-2}
\text{at} \,\, \, \varepsilon^0&:& \quad \gamma_{\varGamma[\varphi_c]} = -\lambda\partial_\lambda C_1(\lambda),
\\
\label{EffP-P-2-2}
\text{at} \,\, \, \varepsilon^{-1}&:& \quad  C_1(\lambda) \gamma_{\varGamma[\varphi_c]} =
-\lambda\partial_\lambda C_2(\lambda) + \beta_4 \partial_\lambda C_1(\lambda),  \quad \text{etc.}
\end{eqnarray}

From one hand, Eqn.~(\ref{EffP-P-2-2}) gives the definition of the ${\rm P}_{\varGamma}$-operator, see Eqn.~(\ref{PR-1}).
At the same time, from the other hand, the pole relations allow us to fix the uncertainties associated with the $\delta(0)$-singularity,
see Eqn.~(\ref{delta-sing}). To demonstrate it, let us first write down the charge and massive terms of $\varGamma[\varphi_c]$,
see Eqn.~(\ref{Delta-G}), in the form of
\begin{eqnarray}
\label{G-m-lambda}
&&\varGamma_{2,4}[\varphi_c] =
  \begin{pmatrix}
  m_0^2 \varphi_c^2 /2 \\
  \mu^{2\varepsilon} \lambda_0 \varphi^4_c/ 4!
  \end{pmatrix}
\Big\{
1+
     \begin{pmatrix}
     d^{\{ m\}}_{(I)}  \\
     d^{\{ \lambda\}}_{(I)}
  \end{pmatrix}
\lambda \, Z_{\lambda} (\lambda)
+ \begin{pmatrix}
     d^{\{ m\}}_{(III,A)}  \\
     d^{\{ \lambda\}}_{(III, A)}
  \end{pmatrix}
\lambda^2 \, Z^2_{\lambda} (\lambda) \,G(1,1)
\nonumber\\
&&
 + \begin{pmatrix}
     d^{\{ m\}}_{(III, B)}  \\
     d^{\{ \lambda\}}_{(III, B)}
  \end{pmatrix}
\lambda^3\,  Z^3_{\lambda} (\lambda) \, G(1,1) G(1, 2-D/2)
\Big\}\frac{\delta(0)}{\Gamma(2)}
\nonumber\\
&&\equiv\begin{pmatrix}
  Z^{-1}_{m}(\lambda) \,\, m_0^2 \varphi_c^2 /2  \\
  Z^{-1}_{\lambda}(\lambda) \,\, \mu^{2\varepsilon} \lambda_0 \varphi^4_c/ 4!
  \end{pmatrix},
\end{eqnarray}
where $d^{\{ m; \lambda \}}_{(i)}$ denote the numerical coefficients associated with the massive and charge terms of the
given diagrams and the charge has been re-expressed via the renormalized quantity
in the diagram contributions which are forming the $Z^{-1}$-factor \cite{Grozin:2005yg}.
We stress that, in contrast to QED/QCD, the renormalization of $\varGamma[\varphi_c]$ is given by the same set of diagrams
We remind that, in QED/QCD, the remormalization of mass and fields are ensured by the two-point Green functions, while the
charge is renormalized with the help of the three-point Green functions etc.

For the sake of shortness, it is convenient to rewrite Eqn.~(\ref{G-m-lambda}) as
\begin{eqnarray}
\label{Rep-1}
\varGamma[\varphi_c] = \sum_{i={(I)}...} \lambda^{n_{(i)}}\, Z^{n_{(i)}}_{\lambda}(\lambda) \, F^{(i)}(\Gamma; \varepsilon)\delta(0),
\end{eqnarray}
where $Z_\lambda$-factor is represented by Eqn.~(\ref{DeltaZ}) and
\begin{eqnarray}
\label{Rep-2}
&&F^{(I)}(\Gamma; \varepsilon)=a_0 + a_1 \varepsilon + a_2 \varepsilon^2 + o(\varepsilon^3),
\\
&&F^{(III,A)}(\Gamma; \varepsilon)=\frac{b_{-1}}{\varepsilon} + b_0 + b_1 \varepsilon + b_2 \varepsilon^2 + o(\varepsilon^3),
\\
&&F^{(III,B)}(\Gamma; \varepsilon)=\frac{c_{-1}}{\varepsilon} + c_0 + c_1 \varepsilon + c_2 \varepsilon^2 + o(\varepsilon^3).
\end{eqnarray}
In Eqn.~(\ref{Rep-1}) we take into account the possibility of dimensional extension for the pre-delta functions mentioned above.

At the order of $[\lambda]^2$, focusing on the $1/\varepsilon^2$- and $1/\varepsilon$-singularities,
the pole relations of Eqns.~(\ref{EffP-P-2}) and (\ref{EffP-P-2-2}) generate the following relation
\begin{eqnarray}
\label{PR-f-1}
C_{22}^{\{\lambda\}}=\big( C^{\{\lambda\}}_{11} \big)^2
\end{eqnarray}
which leads to the relation given by
\begin{eqnarray}
\label{PR-f-2}
a^{\{\lambda\}}_{(III,A)} b_{-1} = \big( a^{\{\lambda\}}_{(I)}\big)^2 a_0^2,
\end{eqnarray}
where $b_1$ and $a_0$ are known from the direct calculations, while $a^{\{\lambda\}}_{(III,A)}$ and $a^{\{\lambda\}}_{(I)}$
have to be determined.
Without loosing the generality, one can normalize the effective action/potential in order to get $a^{\{\lambda\}}_{(I)}=1$ for the
diagram of $I$-type. Hence, from
Eqn.~(\ref{PR-f-2}), the constant $a^{\{\lambda\}}_{(III,A)}$ can be readily fixed.

In the similar way, the pole relations for $Z_m$-factor give
\begin{eqnarray}
\label{PR-f-3}
2 C_{22}^{\{m\}}=\big( C^{\{m\}}_{11} \big)^2 + C^{\{\lambda\}}_{11} C^{\{m\}}_{11}
\end{eqnarray}
and, hence, the uncertainty fixing relation takes the form of
\begin{eqnarray}
\label{PR-f-4}
2 a^{\{m\}}_{(III,A)} b_{-1} = a^{\{m\}}_{(I)}\big( a^{\{\lambda\}}_{(I)} + a^{\{m\}}_{(I)} \big) a_0^2.
\end{eqnarray}
In Eqns.~(\ref{PR-f-4}), the coefficients $a^{\{m\}}_{(i)}$ and $a^{\{\lambda\}}_{(i)}$ have been chosen to be different
ones. However, there is an extra condition which can re-express one coefficient from another.
Based on the stationary method, we have the functional extremum condition as
$\delta \varGamma[\varphi_c]/\delta\varphi_c= 0$ that leads to
$m^2 + \lambda \varphi_c^2/6=0$.
As a result, the coefficients $a^{\{m\}}_{(i)}$ and $a^{\{\lambda\}}_{(i)}$ cannot be independent ones.

As the next step, concentrating on the order of $[\lambda]^3$, from Eqns.~(\ref{EffP-P-2}) and (\ref{EffP-P-2-2}),
we can readily calculate the coefficient giving the anomalous dimension.
We obtain that
\begin{eqnarray}
\label{AD-PR}
C_{12}^{\{\lambda\}} = \frac{3}{7} \, \frac{C_{23}^{\{\lambda\}}}{C_{11}^{\{\lambda\}}}
\end{eqnarray}
which defines also the operation ${\rm P}_{\varGamma}$ of Eqn.~(\ref{PR-1}).

\section{Generation of $V_{z,x}$-operation to the higher orders}
\label{Sec-Vop-2}

In the preceding section,  we demonstrate the $V_{z,x}$-operation applied to the non-local operator Green function
where the standard interaction vertex ${\it (c)}$ has been taken into account up to the order of $[\lambda]^2$ .
In this section, we present the generation of $V_{z,x}$-operation to the higher order of $[\lambda]$.
Let $G^{(n\ge 2)}_{\mathcal{O}}(x_1, x_2; z_1, z_2)$ be the non-local operator Green function corresponding
to the higher order of $[\lambda]$.
Focusing on the singular part of this function, we have
\footnote{We use the following a shortness notation: $(x_i, z_j)=(x_1,x_2 ; z_1, z_2)$.}
\begin{eqnarray}
\label{GF-ho}
G^{(n\ge 2)\, \text{sing.}}_{\mathcal{O}}(x_i; z_j) = \sum_k G^{(n\ge 2)}_{\mathcal{O}}(x_i; z_j | 1/\varepsilon^k)
\Rightarrow \frac{c^G_k}{\varepsilon^k} + \frac{c^G_{k-1}}{\varepsilon^{k-1}} + ... + \frac{c^G_1}{\varepsilon}  + c^G_0 + o^G(\varepsilon).
\end{eqnarray}
In $\varepsilon$-expansion, the prefactor $C^{(n\ge 2)}(D)$  of Eqn.~(\ref{CD})  being the combination of $\Gamma$-functions
\footnote{
The exact form of $C(D)$ depends on the order of $[\lambda]$}
has a form of series as
\begin{eqnarray}
\label{CD-n}
C^{(n\ge 2)}(D) = 1 + o_1(\varepsilon),
\end{eqnarray}
where $o_1(\varepsilon)$ implies the certain series over $\varepsilon$ depending on the order but the exact form of series is irrelevant for
our consideration, see below.

With these, Eqn,~(\ref{Diam-1-2}) for an arbitrary order takes the following form
\begin{eqnarray}
\label{Diam-1-2-n}
\overline{\varGamma}^{(k\ge 2)}[\varphi_c]= \frac{1}{C^{(k\ge 2)}(D)}\,V_{z,x} \Big\{
G^{(n\ge 2)}_{\mathcal{O}}(x_i; z_j) \Big\}
\end{eqnarray}
or, in other words, we have
\begin{eqnarray}
\label{Diam-1-2-n2}
&&\frac{c^{\varGamma}_{k+1}}{\varepsilon^{k+1}} + \frac{c^{\varGamma}_{k}}{\varepsilon^{k}} + ... +
\frac{c^{\varGamma}_1}{\varepsilon}  + c^{\varGamma}_0 + o^{\varGamma}(\varepsilon)=
\nonumber\\
&&
\Big\{ 1 + o_1(\varepsilon) \Big\}
\,V_{z,x} \Big\{
\frac{c^G_k}{\varepsilon^k} + \frac{c^G_{k-1}}{\varepsilon^{k-1}} + ... + \frac{c^G_1}{\varepsilon}  + c^G_0 + o^G(\varepsilon)
\Big\}
\nonumber\\
&&\equiv
\Big\{ 1 + o_1(\varepsilon) \Big\}
\, \Big\{
\frac{c^{VG}_{k+1}}{\varepsilon^{k+1}} + \frac{c^{VG}_{k}}{\varepsilon^{k}} + ... + \frac{c^{VG}_1}{\varepsilon}  + c^{VG}_0 + o^{VG}(\varepsilon)
\Big\}.
\end{eqnarray}
From Eqn.~(\ref{Diam-1-2-n2}), concentrating on the highest singular terms one can see that
\begin{eqnarray}
\label{c2-coeff}
c^{\varGamma}_{k+1} = c^{VG}_{k+1}.
\end{eqnarray}
Of course, such a simple relation is valid only the highest singular terms due to the universal form of Eqn.~(\ref{CD-n}).
For the other singular terms, one needs the exact form of expansion including the finite terms with respect to $\varepsilon$.

If the anomalous dimension of $G^{(n\ge 2)}_{\mathcal{O}}(x_i; z_j)$, {\it i.e.} the coefficient $c^G_1$, is somehow known,
we use the pole relations to transform the coefficient  $c^G_1$ to the coefficient $c^G_k$ at the highest singular term and, then,
we immediately get the highest singular term of $\overline{\varGamma}^{(k\ge 2)}[\varphi_c]$ with the help of $V_{z,x}$-operation,
see Eqn.~(\ref{c2-coeff}).  Afterwards, we again use the pole relations for $\overline{\varGamma}^{(k\ge 2)}[\varphi_c]$ to derive
the coefficient $c^{\varGamma}_1$. That is, we have the following chain of operations:
\begin{eqnarray}
\label{chainN}
c^G_1 \stackrel{\text{P}_G}{\Longrightarrow} c^G_k \stackrel{V_{z,x}}{\Longrightarrow} c^{\varGamma}_{k+1}
 \stackrel{\text{P}_\varGamma}{\Longrightarrow} c^{\varGamma}_{1},
\end{eqnarray}
where $\text{P}_G$ and $\text{P}_\varGamma$ denote the pole relations written for the Green functions and for the effective potential,
respectively.

As a result, we can derive the anomalous dimension for the effective potential provided we know the anomalous dimension
of the corresponding non-local operator Green function. It is important that this procedure is almost algebraical one which
is very useful for the higher order of corrections.

\section{Conclusion}

As a starting point, the generating functional for the scalar $\varphi^4$ has been reformulated in an alternative way where
the massive parameter has been considered as a sort of interaction.
In our approach, the new $V_{z,x}$-operation that transforms the Green function of non-local operator to
the corresponding vacuum integration has been described in detail.
With the help of this operation and the pole relations, the anomalous dimension of effective potential at $\ell$-loop accuracy
can be almost algebraically derived from the known anomalous dimension of the corresponding non-local operator
at $(\ell-1)$-accuracy.
For the effective potential calculations, the special role has been paid for the treatment $\delta(0)$-singularity/uncertainty
within the sequential approach.

\section*{Acknowledgements}

We thank M.~Hnatic, A.~Manashov, S.V.~Mikhailov and L.~Szymanowski for very useful discussions.

\end{document}